\documentclass[superscriptaddress,eqsecnum,amsfonts,showpacs]{revtex4}
\usepackage{epsfig}

\newcommand{\be}{\begin{equation}}
\newcommand{\ee}{\end{equation}}
\newcommand{\bea}{\begin{eqnarray}}
\newcommand{\eea}{\end{eqnarray}}

\newcommand{\nn}{\nonumber}
\newcommand{\ep}{i\epsilon}
\newcommand{\om}{\omega}

\begin{document}

\title{Gauge Technique approximation to the $\pi \gamma$ production and the pion  transition form factor}
\author{V. \v{S}auli}
\affiliation{Department of Theoretical Physics, NPI Rez near Prague, Czech Academy of Sciences}

\begin{abstract}
  The  pion transition form factor  $G(q^2)$   is computed on the entire domain of spacelike and  timelike
  momenta using a quantum field theory  continuum approach.
In analytical continuation of the function $G(Q^2)$ we utilized  the Gauge Technique with the quark propagator
 determined from Minkowski space solution of QCD Dyson-Schwinger equations. 
The scale is set up by the phenomena of dynamical chiral symmetry breaking, which is a striking feature of low energy QCD. 

\end{abstract}
\maketitle


\section{Introduction}

Transition meson form factors carry nontrivial information about the interaction of quarks  inside of mesons.
 Among them, the pion transition form factor $G(Q_1^2,Q_2^2)$  describes   $2\gamma^*\rightarrow \pi$  transition, turns to be  particularly useful  means  testing of   nonperturbative techniques in  QCD.  
 Aforementioned double variable function $G(Q^2=Q_1^2,Q_2^{2})$,  where $Q_i$ are the momenta of the virtual photons, is measured  in more involved process   $e^+e^-\rightarrow  e^+e^-\pi_0$ and kinematically selected  events   provide the single variable pion transition form factor $G(Q^2)$
 in the limit of   nearly on-shell photon $Q_2^{2}=0$.   The  form factor $G(Q^2)$ at the domain of  spacelike momentum $Q^2>0$   has been recently extracted from the data by the Belle Collaboration \cite{Belle},  earlier by the BABAR \cite{bar}, CELLO \cite{cel} and CLEO \cite{cleo} collaborations.
 In the timelike domain of the four momentum $(q^2>0; Q^2\rightarrow - q^2)$, the form factor $G^2(q^2)$ appears linearly in the cross section of the annihilation process $e^+e^-\rightarrow \pi^o\gamma$ with one of the outgoing photon  perfectly   on-shell. Ignoring the interference with other sources  of $3\gamma$,  the  pion  form factor $G(q^2)$ has been extracted form the data collected  by CMD2/SND collaboration  \cite{SND}.

Using   predictions of perturbative QCD \cite{LEBRO1979,LEBR1980,EFRA1980}  the transition pion form factor 
is  absolutely normalized at asymptotic momentum transfer $Q>>m_{\pi}$ with simple functional form: 
\be \label{PT}
G(Q^2)\rightarrow \frac{4\pi^2 f_{\pi}}{Q^2} \, .
\ee

The validity of perturbative QCD  represented here by Eq. (\ref{PT}) has been  experimentally tested at momenta $Q< 6 GeV$.
Owing the known  discrepancy between the data collected by BaBaR and Belle Collaborations, only 
the later being with an agreement  with the Eq.  (\ref{PT}) at  $Q_0^2\simeq 20 {\mbox GeV}^2$, where $Q_0$ is the energy scale 
of perturbative QCD limit.   The BaBar data overestimated the perturbative asymptotic  of Eq. (\ref{PT}), 
provoking  discussion at that time \cite{BMPS2012},\cite{BCT},  that the BaBar  sample of data does not   accurately represent the pion transition from factor
,  including also  arguments coming  from  the prompt use of Dyson-Schwinger equations \cite{HLL}.

  In this paper we contribute toward understanding of the pion transition form factor by using
 the existing  Dyson-Schwinger equations (DSEs) calculation \cite{VHP}, which has the capacity  to provide nonperturbative results
 not only  for the  Euclidean momentum (spacelike) , but on the entire domain of Minkowski space momentum ( we use the Minkowski metric convention $g_{\mu\nu}=diag(1,-1,-1,-1)$, thus $q^2$ is positive for the timelike momentum , while capital $Q$ is used  for the spacelike domain of momentum, where  $q^2<0; Q^2=-q^2 $ and the sign is omitted when writing the argument in the function $G$ as a part of convention ). The DSEs  are traditional tool for such calculations, they have potential to  provide nonperturbative results in infrared momentum domain, where the QCD coupling is strong, while smoothly approaching the perturbative limit when momenta are asymptotically large $Q^2>>\Lambda_{QCD}^2$. Using the formalism of QCD DSEs , chiral symmetry breaking and its connection with meson spectra
 is  well understood today \cite{MR1997,MT1999,EACK2008,VS2014,HGKL2017,DKHK2014}. The  electromagnetic form-factors of pion and kaon \cite{MT99,BPT2003,kacka2005,CHA2013} and meson transition form factors \cite{RCBCGR2016,WEFW2017,EFWW2017,DRBBCCR2019} were studied in unified picture of Euclidean space QCD DSEs.

The development of nonperturbative tools which incorporates confinement and  naturally avoid the use of  on-shell quark spinors
is one of  the aims of presented paper. A confining solution for the quark propagator as obtained in minimal symmetry preserving approximation of QCD DSEs studied in the paper \cite{VHP} is used to construct two photon gauge vertex. For this purpose we use the  Gauge Technique developed in \cite{salam63,saldel1964,strath1964,DEWE1977,DEZH1984} and  employed in practice in \cite{Cor1982,ARBR2003,Sa2012,CHA2013,VHP}.

The paper is organized as the following:
Next two sections describe the equations needed for 
calculation of the pion transition form factor. Particularly, the Gauge Technique is reviewed in the Section III. 
 The main result, the integral expression relating the quark propagator with the pion transition
 amplitude is presented in the Section IV. Numerical results are presented in the Section V and we  conclude in  Section VI.
Some subtleties relating with the extraction of spectral representation from the quark gap equation is 
described in the Appendix.

\section{Equations for transition form factor}

 The two photons pion transition,  which determines the function $G(q^2)$  is computed from
 \be
 M_{\mu\nu}(k,l)= T_{\mu\nu}(k,l)+ T_{\nu\mu}(k,l) \,\, ,
 \ee
where the pion's momentum $P=k+l$, $k$ and $l$ are the photon momenta and two variable form factor is defined 
as
\bea 
T_{\mu\nu}(k,l)&=&\frac{e^2}{4\pi^2}\epsilon_{\mu\nu\alpha\beta}k^{\alpha}l^{\beta}G(k^2,l^2) \,  ,
 \label{udef} \\
M_{\mu\nu}(k,l)&=&i\int \frac{d^4 q}{(2\pi)^4} \Gamma_{\pi}(q_1,q_2)G_{\mu\nu}(q_1,q_2;k,l) \, ,
\label{tykev}
\eea
where $  \Gamma_{\pi}(q_1,q_2)$ is the pion Bethe-Salpeter vertex function for which purpose we use the solution of  the pionic Bethe-Salpeter equation in ladder-rainbow approximation as obtained in \cite{VHP}. The model is  described in the Appendix for purpose of completeness.   

The Two Lorentz indices function $ G_{\mu\nu}(q_1,q_2;k,l)$ in the Eq. (\ref{tykev})
is non-amputated fully dressed quark-antiquark-two gamma vertex function in notation where variable $q_1$ and $q_2$ stands for the constituent (anti)quark momenta.  The function $ G_{\mu\nu}$ consistent with rainbow-ladder approximation can be written
 uniquely as the  decomposition of the quark propagators $S$ and the proper  quark-photon vertices
 $\Gamma$ 
 
\be \label{mnauk}
 G_{\mu\nu}(q_1,q_2;k,l)=S(q_1)\Gamma_{\mu}(q_1,q_1-k)S(q)\Gamma_{\nu}(q_2-k,q_2)S(q_2)
+(k\rightarrow l) \, ,
\ee
with momenta labeling that keeps $q_1=q_2+k+l$. Putting the Eq. (\ref{mnauk}) into (\ref{tykev})
one gets  the integral over the three quark propagators with three vertices placed among them. Diagrammatically thus they consist the well-known  triangle diagram, which represents a common starting approximation  \cite{RCBCGR2016,DRBBCCR2019} .

To get a single variable pion transition form factor, the kinematics constrains read
\be 
  l^2=0 , \, \, P^2=k^2+2k\dot l=m_{\pi}^2\, \, ,
\ee
albeit we will keep both photons off-shell during the derivation here.

In order to express $G_{\mu\nu}$  we employ Gauge Technique  where we 
use  confining solution for the  quark propagator as obtained in  $S$ \cite{VHP}. 
Recall, the  obtained  quark propagator is known in the entire Minkowski space
and  the absence of  on-shell pole is a striking and important  property of the solution.  
 
Perturbative and nonperturbative parts unavoidably matches in any formula 
for  exclusive process in QCD. 
At this point remind here the well known QCD  factorization formula for the transition form factor, which has led to the Eq.(\ref{PT}). 
It has the following structure:
\be 
G(Q^2)=4\pi^2 f_{\pi}\int_0^1 dx T_H(x,Q^2,\alpha(\xi),\xi)\phi_{\pi}(x,\xi) \, .
\ee
It contains nonperturbative part: the pion distribution function $\phi$, which can be expressed as the light cone projection of 
Bethe-Salpeter amplitude \cite{LEBRO1979}, \cite{RCBCGR2016} and the hard part of the parton  scattering $T$ is evaluated perturbatively 
within the use of on-shell quark spinors. 

On the other hand,  the  phenomena of confinement in QCD can be mathematically formulated to the statement 
that  there are no free on-shell quarks in the Nature. In other words,
  the  on-shell spinors could not  be an adequate description for the confined quark modes inside  hadrons.
  In this respect, we  elucidate the sensitivity of the
 $\gamma^*\gamma \to \pi^0$ transition form factor to the  behavior of the quark propagator $S$ in presence of  analytical confinement.


\section{Gauge Technique}
\label{gaugen}

The proliferation of Nakanishi's Perturbation Theory Integral Representation from scalar theory  \cite{NAKAN} to vertices including fermions began with the paper \cite{SAULI}. Note, the generalization to the abelian quark-photon vertex attracted attention already more than  two decades ago \cite{Del1999}.  Since till today, the  numerical realization of such a program is still not yet completed, we begin with the Gauge Technique approximation, 
 which presumably will be a part of more sophisticated but not yet completed integral representation for this vertex. 

 The Gauge Technique is a well-established method \cite{salam63,saldel1964,strath1964,DEWE1977,DEZH1984,Del1999} of solving Dyson-Schwinger equations in gauge theories. The main idea is construct an Ansatz that expresses a given vertex as functional of the propagator spectral function, such that the corresponding Ward-Takahashi \cite{WA1950},\cite{GR1953},\cite{TA1957} or Slavnov-Taylor identity is automatically satisfied. In our case, the Gauge Technique consists of writing a  solution  for  the  double vector  un-amputated  vertex  in  the following  form
\bea \label{double}
G^{\mu\nu}(q_1,q_2;k,l)=\int_{\Gamma} d x \rho(x) \frac{1}{\not q_1-x}[\gamma^{\mu}\frac{1}{\not q_1-\not k-x}\gamma^{\nu} 
\nn \\
+\gamma^{\nu}\frac{1}{\not q_2+\not k-x}\gamma^{\mu}]\frac{1}{\not q_2-x} \, ,
\eea
 where we used a generalized spectral representation
for the quark propagator
\be \label{spectral}
S(p)=\int_{\Gamma} d x \frac{\rho(x)}{(\not p-x)} \, .
\ee
 
 Actually it is easy to show, that $G^{\mu\nu}$ given by the Eq.(\ref{double})  satisfies the Ward-Takahashi identity:
 \be  \label{verca}
k_{\mu}G^{\mu\nu}(p^{,},p;k,l)=G_{\nu}(p^{,},p+k)-G_{\nu}(p^{,}-k,p) \, ,
\ee
where
\be  \label{delbourg}
G^{\mu}(p,q)=\int_{\Gamma} d x \rho(x) \frac{1}{\not p-x}\gamma_{\mu}\frac{1}{\not q-x} \, , 
\ee
which is the one body form of Gauge Technique approximation to the dressed (untruncated) quark-photon vertex $(G=S\Gamma S)$.
 This is established again  that the vertex (\ref{delbourg}) satisfies Ward-Takahashi identity:
 \be
 (p+q)_{\mu}G^{\mu}(p,q)=S(p)-S(q) \, ,
\ee
for propagator satisfying (\ref{spectral}).
 We use the function $\rho$  as obtained in the paper \cite{VHP}  and  we describe 
 some details of   its extraction in the Appendix here.

The main drawback of aforementioned approach is that it leaves  the transverse part of the vertex undetermined.
 Attempts   to solve the problem have been done  by using a wider set of the Euclidean space DSEs  \cite{MT99,ROSS2000,MISA2019,will18} or  by resorting  the transverse Ward identities \cite{HE2001,HEKHA2006,PENWIL2006,tran4,tran5,tran6}.

\section{Method of evaluation}

To make the rainbow-ladder  truncation of DSEs a  meaningful approximation of QCD  a  common practice used for instance in  \cite{MR1997,MT1999,EACK2008,VS2014,HPRG2015,HGK2015,AW2015,HGKL2017,VHP,RW2019,QRS2019} is to solve the quark gap equations
 simultaneously within the Bethe-Salpeter equation
for  meson.   The same was  done for model used herein in the paper \cite{VHP} for the first time and fitting procedure was just repeated for our purpose here. The  model  parameters  were varied in order to meet pionic   observable: the pion mass and pionic weak decay constant.
  For this purpose  the pion bound state vertex function was obtained via solution of the  Bethe-Salpeter equation:


%
 \bea \label{BSE}
\Gamma_{\pi}(P,p)&=&i  \int\frac{d^4k}{(2\pi)^4}\gamma_{\mu}S_q(k_+)\Gamma_{\pi}(P,k)S_q(k_-)\gamma_{\nu} [-g^{\mu\nu} V_v(p-k)-4/3\xi g^2\frac{L^{\mu\nu}(p-k)}{(p-k)^2}] \, ,
\eea
 where for the sake of consistency, 
the Bethe-Salpeter  and the quark Dyson-Schwinger equation must have the identical kernel. Here we use the model \cite{VHP}, which 
has simple functional form 
\bea \label{potent}
V_{v}(q)&=&\int d \om \frac{\rho (\om)}{q^2-\om+\ep}
\nn  \\
\rho_v(\om)&=&c_v [\delta(\omega- m_g^2)-\delta(\omega-m_L^2)] \, \, ,
\eea
with the product of $SU(3)$ color matrices $\lambda_a\lambda_a$ absorbed in the constant $c_v$ and where $L^{\mu\nu}$ stands for the longitudinal projector $L^{\mu\nu}(q)=q^{\mu}q^{\nu}/{q^2}$.

In the Eq. (\ref{BSE}) $P$ is the total momentum of meson satisfying $P^2=M^2$, $M=140 {\mbox MeV} $ for the ground state and  the arguments in the quark propagator are $k_{\pm}=k\pm{P/2}$.  The DSE/BSE system provides the solution    with   small error 
 $h\simeq 5.0\,  10^{-5}$ defined by ( \ref{norms}) for the quark propagator calculated in  the  gauge   $(g\xi)^2/(4\pi)^2=0.18$  and the  kernel couplings  (\ref{potent}) $c_v/(4\pi)^2=-1.8$ and   $m_g^2/m_L^2=2/7.5$ where  $m_g$ in  physical units is  $m_g=556 {\mbox MeV} $. 
 
 The pion  BSE vertex  function  reads  
\bea
\Gamma_{\pi}(P,p)=\gamma_5\left(\Gamma_A(P,p)+\not p \Gamma_B(P,p)+\not P \Gamma_C(P,p)+[\not p,\not P] \Gamma_D(P,p)\right) \, ,
\eea
where all  $\Gamma_X$ was used to determine the pion mass and where  we skip the isospin matrix notation,
 which is not necessary in given truncation of DSE system.  The evaluation of BSE and decay constant was performed in  
standard manner and reader can find an appropriate  eigenvalue method solution   in  \cite{VS2014} and $f_{\pi}$ was calculated according 
to Eq.(33) in   \cite{MR1997}, representing  codes used for purpose of this study are available at \cite{gemma}.

The gauge-independence of physical quantities is essentially assumed, and we have used   a nonzero gauge fixing parameter $\xi$ to  accelerate numerical convergence of  equations. Due to the  required stability of all equations   a single gauge  $(g\xi)^2/(4\pi)^2=0.18$ was finally chosen at the end of  iteration process . The numerical  convergence is not  better  for presented model in Landau gauge (the gauge  almost exclusively chosen in other studies).
 At this stage we do not attempt to search for solutions  at other individual  gauges and leave this task for future  exercise.

 In what follows we describe approximations made that allows to write the pion transition entirely in terms of the quark propagator. First of all, let us note that 
  the production of the vector strangeonium represented  by    $\phi$ meson peak is suppressed by  two orders when compared  the $\omega/\rho $ peak in experimental $\pi\gamma$ cross section  \cite{SND} and therefore  we restrict to two flavors $f=u,d$ in its isospin (equal mass) limit.

In the chiral limit , which we subsequently employ exclusively,
the following  Goldberger-Treiman-like identity \cite{98,ICC2013,ICC2014}:
\be \label{GTI}
\Gamma_A(0,p)=\frac{B(p)}{f_\pi}\, 
\ee
 exhibits the  equivalence between the one-body quark and pseudoscalar two-body problem in QCD.

The renormalized quark function $B$ obtained from the Eq. (\ref{gap}) satisfies the dispersion relation
\be
B(p)=m_q+\int do \frac {\rho_B(o)}{p^2-o+\ep}=m_q(\zeta)+\int do \frac {\rho_B(o)}{p^2-o+\ep}-
\int do \frac {\rho_B(o)}{\zeta-o+\ep} \, ,
\ee
where $\rho_B$ is the imaginary par of selfenergy $\Sigma_B$  and can be easily extracted form the solution of DSE.   
In this study such   we replace the pion vertex function in the expression for transition form factor (\ref{tykev}) 
by what  $B$ could be in the chiral limit 
\be  \label{gold}
\Gamma_{\pi}(p,P)=\gamma_5 \frac{1}{\cal {N}}\int do \frac {\rho_B(o)}{p^2-o+\ep} \, ,
\ee
where ${\cal{N}}$ is the normalization of the BSE vertex, satisfying approximately ${\cal {N}}=f_{\pi}$, with its exact value dictated by  
 the canonical normalization of the BSE vertex \cite{LS1969}.

Substituting (\ref{gold}) and gauge technique integral representation  into (\ref{tykev} , one gets  the following expression for the double variable pion transition form factor $\epsilon_{\mu\nu\alpha\beta}k^{\alpha}l^{\beta}G(k^2,l^2)\simeq I_{\mu\nu}(k,l)$ 
\bea \label{trik}
I_{\mu\nu}(k,l)&=&Tr \int d \om  \int_{\Gamma} d\sqrt{s} \int\frac{d^4p}{(2\pi)^4} \frac{\gamma_5 \rho_B(\om)}{p^2-\om+\ep}\frac{\not p+\not{P}/2+\sqrt{s}}{(p+P/2)^2-s+\ep}\left[
\gamma_{\mu}\frac{\not p+(\not k -\not l) /2+\sqrt{s}}{(p-\frac{k-l}{2})^2-s+\ep}\gamma_{\nu}\right.
\nn \\
&+&\left.\gamma_{\nu}\frac{\not p+(\not l -\not k) /2+\sqrt{s}}{(p-\frac{l-k}{2})^2-s+\ep}\gamma_{\mu}\right]
\frac{\not p-\not{P}/2+\sqrt{s}}{(p-P/2)^2-s+\ep} \rho(\sqrt{s}) \, \, ;
\eea
where the  integral over the variable $\omega$  is due to the Eq. (\ref{gold}) and the integral over the variable $\sqrt{s}$ is due to the Eq. (\ref{double}) .
Two weight functions $\rho$ and $ \rho_B(\om)= -\Im B(\om)/\pi $ are not independent, since related through the definition  (\ref{cuba}).

In what follow we integrate over the momentum analytically and   within the help (\ref{spectral}) we rewrite the integral  (\ref{trik}) equivalently 
into the expression where we integrate over the known and regular function: the  quark propagator.  
  For this purpose  we  match all denominators in Eq. (\ref{trik}) by using the Feynman paramaterization. Let us  use the variable $x,y$ to match the  quark propagators denominator as expressed in spectral representations. 
 Then, using the Feynman variable $z$,  the obtained result we further match  together with  $p^2-\om$ in the denominator. After a usual shift of the momentum, which complete the square, we perform the  Wick rotation  and momentum integration.  We get the following result  (again in 
Minkowski space):
\bea \label{pinok}
I_{\mu\nu}&=&\frac{1}{(4\pi)^2}\int d \om \rho_B(\om ) \int_{\Gamma} d\sqrt{s} \sqrt{s} \rho(\sqrt{s}) \int_0^1 dx dy dz \frac{-i4 \epsilon_{\alpha\beta\mu\nu} k^{\alpha}l^{\beta}y z^2}{\left[f(k,l;x,y,z)-s z- \om(1-z)\right]^2}   
- k\rightarrow l \, ,
\nn \\
f(k,l;x,y,z)&=&\frac{P^2}{4} yz +r^2(1-y)z-z^2\left[\frac{P^2}{4} (1-2x^2)y^2+\frac{P.r}{2}(1-2x)y(1-y)+\frac{r^2}{4}(1-y)^2\right] \, ,
\eea
 where we have  labeled $r=k-l$. After factorization of the variable  $z^2$ out of the numerator, then according to (\ref{spectral}),  one can recognize that the integrand together with the  factor
  $\int d\sqrt{s} {s}$ is once differentiated quark propagator,  which is  further integrated over the auxiliary variables $x,y,z$.
 The expression (\ref{pinok}) is exact and  if needed  it can be  straightforwardly used  in order to get the result for the pion transition form factor. 
 
In order to reduce number of auxiliary integrals,  one can get very accurate expression by  neglecting  the small term 
$z^2m_{\pi} ^2(1-2x)^2 y^2/4$, which includes the pion mass   in the  denominator of Eq. (\ref{pinok}). 
Doing this explicitly and integrating over the variable $x$ one gets
\bea 
I_{\mu\nu}&=&\frac{1}{(4\pi)^2}\Sigma_{\pm}\int d \om \rho_B(\om ) \int_{\Gamma} d\sqrt{s} \sqrt{s} \rho(\sqrt{s}) \int_0^1  dy dz 
\frac{-i4 \epsilon_{\alpha\beta\mu\nu} k^{\alpha}l^{\beta}}{(1-y) z\left[t_{\pm}-s+\ep\right]}   
- k\rightarrow l
\nn \\
&=&\frac{1}{(4\pi)^2}\Sigma_{\pm}\int d \om \rho_B(\om ) \int_0^1  dy dz 
\frac{-i4 \epsilon_{\alpha\beta\mu\nu} k^{\alpha}l^{\beta}}{(1-y) z}S_s(t_{\pm})   
- k\rightarrow l
\nn \\
t_{\pm}&=&P^2/4 y +r^2(1-y)-z\left(\pm\frac{P.r}{2}y(1-y)+\frac{r^2}{4}(1-y)^2\right)-\frac{\om(1-z)}{z}\, ,
\eea
where the variable $\sqrt{s}$ was absorbed into  the usual  definition of the scalar part of the quark propagator (\ref{cuba}).

Taking $l^2=0$ , the pion transition form factor finally reads
\bea
G(k^2)&=&F(k,l)+F(l,k)
\nn \\
F(k,l)&=&\frac{const.}{(4\pi)^2}\Sigma_{\pm}\int d \om \rho_B(\om )  \int_0^1  \frac{ dy dz}{ (1-y) z} S_s(t_{\pm})    \, ,
\label{numero}
\eea
where one has to take $P.r=k^2$ and $r^2=k^2$ in the function $F(k,l)$ in the expression (\ref{numero}) and 
one should take in mind that  $P.r=-k^2$, when evaluating the function $F(l,k)$.

 \section{Discussion of Solution}
 
 We begin with  discussion  of the spacelike  asymptotic behavior of the pion transition form factor 
$G(Q^2) $, which has curious dependence on the interaction between quarks. 
 It was shown in \cite{HLL},  that in the case of   
 the contact interaction typical for quarks in Nambu-Jona-Lasinio models,   the asymptotic o f$G(Q^2) $  differs from   (\ref{PT}) such that
\be  \label{hll}
G(Q^2)\simeq \frac{2M^2}{Q^2}\ln^2(Q/M) \, ,
\ee
where $M$ is constant constituent quark mass obtained  within  the contact interaction.

Interesting asymptotic form was obtained  \cite{KK1999}  for the double variable transition form factor in  the symmetric limit 
\be  
\label{mef}
Q^2G(Q^2,Q^2)\simeq\frac{2}{3}Q^2G(Q^2,0)= \frac{2}{3}Q^2G(Q^2) \,   ,
\ee   
which tells us, that up to the  prefactor, the same asymptotic behavior persist  for the double variable transition form factor as well.

\begin{figure}
\centerline{\includegraphics[width=9.0cm]{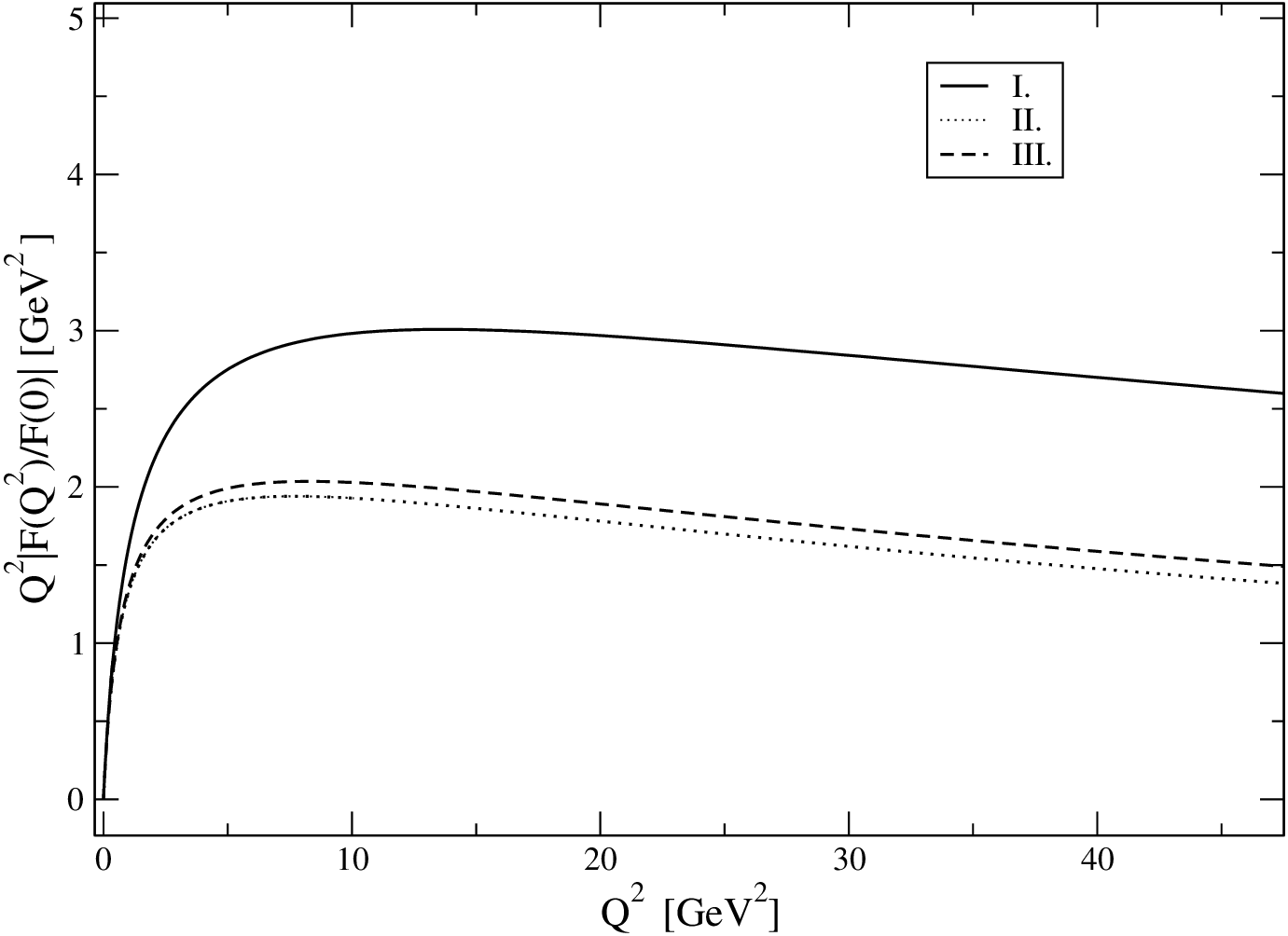}}
\caption{Pion transition form factor $G(Q)$ for the  spacelike momentum  $Q$ calculated as  described in the text: the quantity ($ Q^2G(Q^2)/G(0)$) is shown.
 Each  line labeled by I  II and III use a different interpolator for the pion vertex function described in the text.  }
\label{transpace1}
{\mbox{-------------------------------------------------------------------------------------}}
\
\end{figure}

 From the equations  (\ref{PT}), (\ref{hll}) a specific logarithmic dependence of $Q^2G(Q^2)$ seems to be governed by modeled interaction as follows:
 a weaker interaction between the quarks at  large relative  momentum a smaller power of log (anomalous dimension)  one gets.
 In our simple model, the interquark interaction consists from the  interaction softer then QCD and from the purely longitudinal  term.  
 The solution is shown in the Fig. \ref{transpace1},
where the function $Q^2G(Q^2)$ is displayed.
As we can see, the presence of longitudinal mode is ignored and  the expected behavior  is confirmed  here. The pion transition form factor does not reach its QCD perturbative asymptotic, but instead, the function $Q^2G(Q^2)$ shows up the maximum at few GeV, where it starts to decrease with negative anomalous dimension.

\begin{figure}
\centerline{\includegraphics[width=9.0cm]{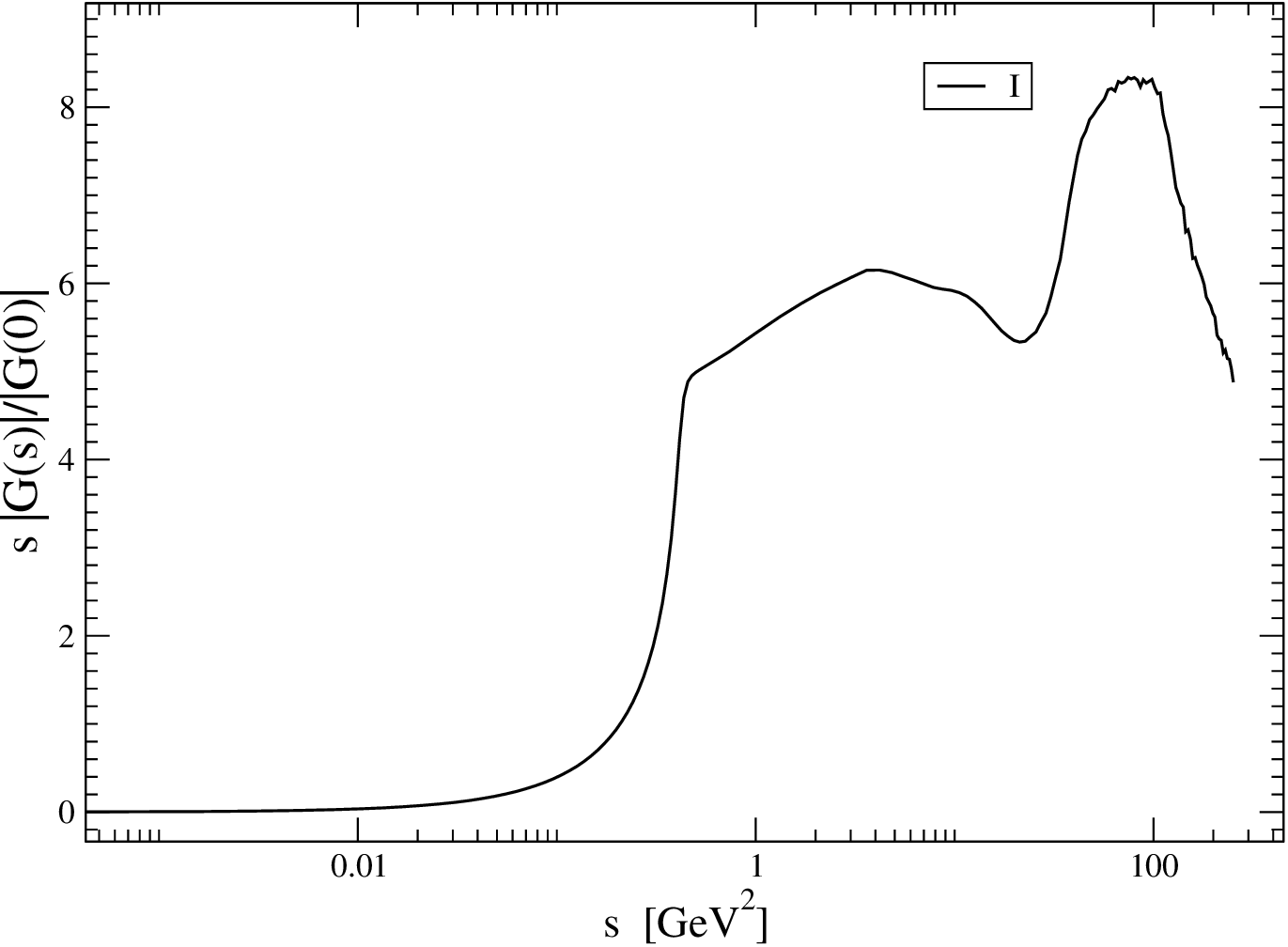}}
\caption{Pion transition form factor $G(q^2)$  for the  timelike argument  $s=q^2$ as described in the text: the quantity ($ q^2G(q^2)/G(0)$) is shown.  }
\label{transitime2}
{\mbox{-------------------------------------------------------------------------------------}}
\
\end{figure}

Similar asymptotic form is observed  for the large timelike  momenta as well, however it happens at  much larger asymptotic momentum $q_o\simeq 10 {\mbox GeV}$ as is shown in the Fig. \ref{transitime2}. 
Before the asymptotic is reached, the function   $q^2G(q)\rightarrow const $ reach the flat plateau with its  detailed shape arises due 
to the interference of all components entering the form factor.

\begin{figure}
\centerline{\includegraphics[width=9.0cm]{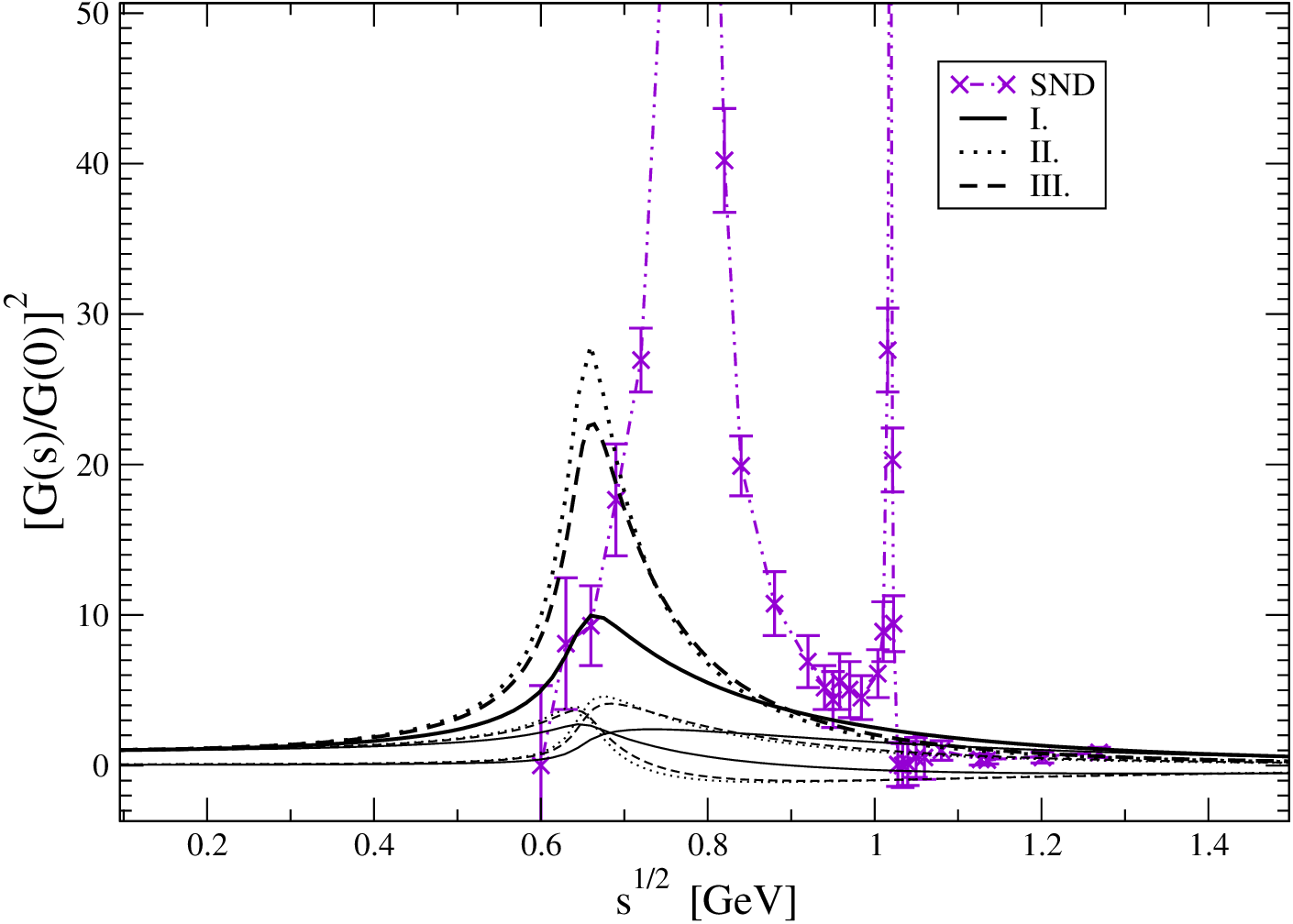}}
\caption{Square of the pion transition form factor $G(s)$ (thick) and $\Re$ and $\Im$ parts of $ G(s)/G(0)$ (thin lines)  shown for the  timelike argument  $s$ and compared to SND data. }
\label{transitime}
{\mbox{-------------------------------------------------------------------------------------}}
\
\end{figure}

 We also show  the graph  of square of the pion transition  form factor  against 
the energy $\sqrt{s}$ in the Fig. \ref{transitime}, and compare to the extracted data from the SND cross section $\sigma_{\pi,\gamma}$.
Considering  the fact that SND data are far from  the threshold $E_{th}= m_{\pi}$ and the properly normalized
ratio $G(q)/G(0)$ is  extracted from the experiments by using:
\be
G(q^2)/G(0)=\frac{\sigma_{\pi,\gamma} 12 \pi^2 f_{\pi}^2}{\alpha^3_{QED}}
\ee
with $\alpha_{QED}$ being the fine structure constant and $\sigma_{\pi,\gamma}$ is the total cross section of 
the process $e^+e^-\rightarrow\pi^0\gamma$. In Fig. \ref{transitime} we also show the imaginary and the real parts of form factors. 
To  guide the reader  eyes, note that those lines optically vanishing at zero momenta stand for the imaginary parts.   

A single line labeled by ``I'' corresponds to the form factor as obtained from the Eq. (\ref{numero}) in all presented figures.  
There is a cusp- the remnant of standard threshold enhancement-  located at $\sqrt(s)\simeq 500 {\mbox MeV}$ .
 The position of this cusp  is dictated by the shape of the quark propagator and  it appears at value $2M(0)$, where $M(0)$
 is the quark constituent mass. We expect the cusp effect could be  even more pronounced for heavier quarks
 due to the smaller washout of perturbative pole in the quark propagator.
 More generally, the  quark quasi-threshold cusp must be a universal part of  many other  measured cross sections.

Since large enough, observed theoretical cusp should be already part of observed structure in $\pi\gamma$ cross section.   
 In case studied it should be located near of inside the  peak of $\omega$ and $\rho$ resonances. 
 Since the obtained cusp is obtained bellow $\rho$ meson mass, 
 it  is likely that the amount of generated dynamical mass is slightly underestimated in our simple model.
 When  rescaling   all dimension-full quantities in  (\ref{potent}) by hand, such that we move  the position of the  cusp into the $\rho-\omega$  mesons centered mass, then  instead of  $140 {\mbox MeV}$ light pion one gets   $m_{\pi}=200{\mbox MeV}$ afterwords. In other words, the pion  gets  heavier compared to other scales in theory, eg. to the vector meson masses, which is likely what happen in our Schwinger-Dyson equations model. We therefore expect the cusp should move into higher scale in a more realistic  model (say, with a more realistic interacting kernel, if one does need to go beyond LRA at all).

Suggestion has been raised \cite{ARBR2003}, that the Gauge Technique alone can provide more significant portion of the $\rho-
\omega$ peak. This claim is unsupported by more self-consistent  approach presented here. Actually,   even if we would use the quark  propagator with the real pole presented in and   enhance the  pseudothreshold cusp into the ordinary threshold, than its absolute  value is far not enough to describe entire experimental $\rho$ meson peak. Some additional important transverse pieces of the quark-photon vertex are needed for a more correct description.

To see other possible effects, we have  checked   changes in $G(q^2)$ behaviour, which stem from different extrapolations  used for the pion Bethe-Salpeter vertex function.   For this purpose we consider the  interpolator  suggested in \cite{ICC2013}, which  takes the form    
\be   \label{cina}
\Gamma_{\pi}(p,P)=\gamma_5 \frac{1}{\cal {N}}\int_{-1}^{1} du \frac {\rho_5(z)}{(k+u P^2/2)^2-\Lambda_{\pi}^2+\ep} \, .
\ee

Furthermore,  we have used more general two variable integral representation for the pion vertex function:
\be   \label{cina2}
\Gamma_{\pi}(p,P)=\gamma_5 \frac{1}{\cal {N}}\int d \om \int_{-1}^{1} dz \frac {\rho_5(\om,u)}{(k+u P^2/2)^2-\om+\ep} \, ,
\ee
which arises when one studies the integral representation for the  pseudoscalar  BSE in the Minkowski space  self-consistently \cite{SAULI} .

Then our derived formula for the pion transition form factor can be written as:
\bea \label{umero}
G(k^2,l^2)&=&F(k,l)+F(l,k) \, \, ;
\nn \\
F(k,l)&=&const. \int d \om du \rho_5(\om,u )  \int_0^1  \frac{ dy dz}{ (1-y) z P.r} \left[S_s(p^2_+)-S_s(p^2_-)\right]    \, \, ;
 \\
p^2_{\pm}&=&r^2(1-y)+\frac{r^2}{4}(1-y)^2 z+\frac{P^2}{4}y+\frac{P^2}{4}u^2\frac{1-z}{z}
\nn \\
&+&\om\frac{1-z}{z}-\frac{P.r}{2}(1-y)\left[\pm y z+(1-z)u\right] \, \, 
\eea
valid for both cases (\ref{cina}) as well as (\ref{cina2}). Note that the  representation  (\ref{cina}) is a special case of the second  (\ref{cina2}), where simplified version of  the Bethe-Salpeter weight function includes  the delta function $\delta(u-\Lambda_{\pi})$ as an assumption.

\begin{figure}
\centerline{\includegraphics[width=9.0cm]{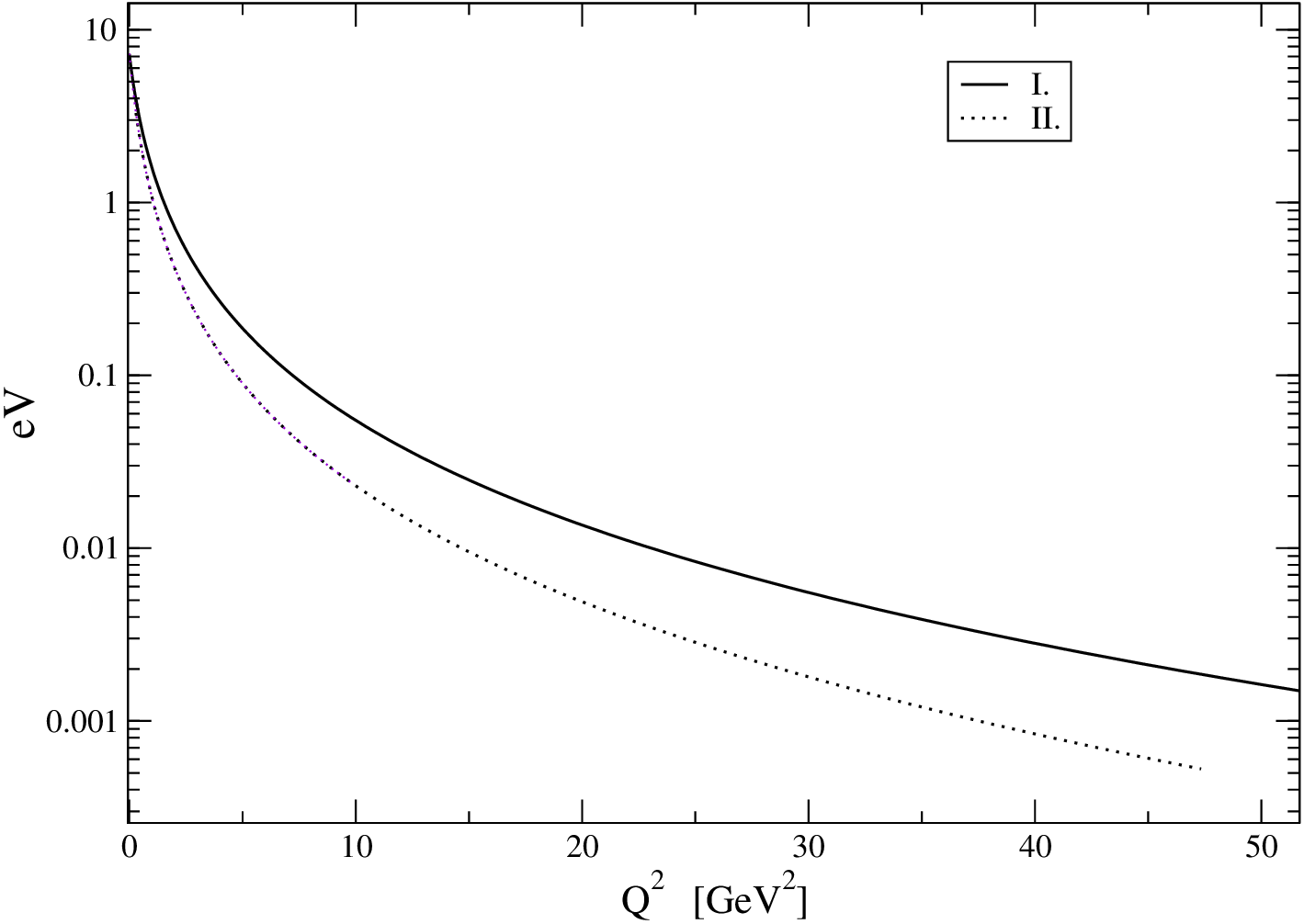}}
\caption{Pion transition form factor $G(Q)$ for spacelike $Q$ as described in the text: the quantity $\Gamma_{\gamma\gamma}(Q)=(\pi/4)\alpha_{QED} m^3_{\pi} G^2(Q^2)$ is shown. Form factors are normalized such that the pion width is reproduced
$\Gamma_{\gamma\gamma}(0)=7.23 {\mbox eV}$ . }
\label{transpace2}
{\mbox{-------------------------------------------------------------------------------------}}
\
\end{figure}

The results for the choice $\rho_5(z)=(1-z^2)$ are labeled as the model II. and added in figures for comparison.
Obviously a  large  enhancement of the cusp in the timelike region is the main feature of presented approximation.
The approximation made  does not affect the decreasing character 
of the  asymptotic behavior  in the  ultraviolet spacelike as seen  figures  \ref{transpace1}  and \ref{transpace2}. A similar statement  is valid for behaviour of $G$  in the timelike domain  of momenta, which can be seen in  the Fig. \ref{transitime2}. 

 It is interesting that almost the same pion form factor $G(Q^2)$ is achieved  with  more primitive
interpolator:
\be   \label{cina3}
\Gamma_{\pi}(p,P)=\gamma_5 \frac{1}{\cal {N}} \frac {1}{k^2-\Lambda_{\pi}^2+\ep} \, ,
\ee
for  which the formula (\ref{umero}) applies as well.
 The lines which represent the pion transition  form factor evaluated with  use of  the interpolator (\ref{cina3}) are labeled
 by III in all presented figures.
For better comparison of all form factors, the constant $\Lambda_{\pi}$ is taken as $\Lambda_{\pi}=m_g$ and  we have used rescaled quark propagator as discussed above. 

As seen in the Fig.\ref{transitime} the cusps are similarly and largely pronounced for  interpolators II (\ref{cina2}) and III (\ref{cina3}) , mainly due to the fact that the  pion vertex has similar unphysical singularity  in both cases. Even so, all obtained results are far from  revealing  structure of vector resonances observed in the experiment, which further support the aforementioned  above: to reveal the form factor structure at resonances, one needs to incorporate transverse vertices.   

The pion transition form factor  $G(Q^2)$ in approximation  I. is compared to the approx.  II.   in the Fig. (\ref{transpace2}). 
The difference is attributed to a different asymptotic of approximated Bethe-Salpeter pion vertex function.
 In accordance to observation made in  \cite{EFWW2017}, a softer interaction provides anomalously decreasing form factor for a large $Q$. 
In this analysis  we do not optimize the model by  further improving  the pion vertex function fit and we let the error freely propagate to higher $Q$.

\
\

\

\section{Conclusions and Prospects}

We completed a computation of the pion transition form factor  $G(q^2)$ in two flavor QCD. 
Theoretical Nakanishi paramaterization for  correlators allows to compute $G(q^2)$ 
on the entire domain of spacelike as well as of timelike momenta in Minkowski space. 
The later represents the analytical continuation of the former  and  numerical evaluation of the form factor $G$ 
for the timelike arguments appears for the first time  in the literature.   
All required elements e.g. the quark propagator within its  spectral function are 
determined by the solution of QCD's Dyson-Schwinger equations obtained in the rainbow-ladder truncation, 
 the leading order in a systematic and symmetry preserving approximation scheme.  
 New developed  technique based on the subtraction at the timelike domain of momenta 
 was the important step to accomplish desired analytical continuation of QCD quark propagator.

A cusp observed at square of momentum $q^2\simeq 4(B(0))^2$ is a universally  predicted as quasithreshold of confined quarks.
 Observed cusp  in the square of the pion transition form factor is about one or two order of magnitudes smaller 
 then $\omega/\rho$ peak observed in the experiment \cite{SND}.   A missing amount of the function  $G(q^2)$ in resonance region needs  an inclusion of important transverse components form factors in the quark-photon vertex.   The formalism developed herein represents small step towards our better understanding of how  confinement mechanism can be  incorporated in unified picture of hadron production form factors in the entire Minkowski space.
 
 \
 
 \

{\bf Acknowledgments}
I am grateful to Petr Byd\v{z}ovsk\'{y} and Daniel Gazda for their help with the manuscript.

\appendix
\section{Quark gap equation in Minkowski space-method and properties of solution}  

  Only very recently, an easy way  to get rid of resisting numeric   was  recognized in \cite{VHP}.  
Since the knowledge how to obtain  spectral functions in a nonperturbative framework of Dyson-Schwinger equations  does not belong to a conventional  wisdom, we describe the method  in details in this Appendix. 

Neglecting the weak interaction,  the  quark propagator  is completely characterized by   two  scalar functions $A$ and $B$ 
(or alernatively $S_v$ and $S_s$) : 
\be \label{cuba}
 S(q)=\not q S_v(q)+ 1 S_s(q)=[A(q)\not q-B(q)]^{-1} \, \, ,
\ee
where the renormalization scheme independent  dynamical mass function is conventionally defined as $M=B/A$.
The quark propagator is obtained from the solution of Dyson-Schwinger equation
\bea   \label{gap}
S^{-1}(q)&=&\not q- m_q-\Sigma(q) \, ,
\nn \\
\Sigma(q)&=&i\int\frac{d^4 k}{(2\pi)^4} \frac{\lambda^a}{2}\gamma_{\mu} S(k) \frac{\lambda^a}{2}\Gamma^{\nu}_{qg} (k,q) D_{\mu\nu}(k-q) 
 \label{munk}
 \eea
where  the product of the gluon propagator $D$ and the quark-gluon vertex $\Gamma_{qg}$ in the  exact expression (\ref{munk} ) was approximated by  the interaction kernel \ref{potent}, such that the selfenergy is
 \bea 
\Sigma(k) &=&-i\int\frac{d^4 k}{(2\pi)^4} \gamma_{\mu} S(k) \gamma^{\mu} V_v(k-q) 
\nn \\
&-&i\frac{4\xi g^2}{3}\int\frac{d^4 k}{(2\pi)^4} \gamma_{\mu} S(k) \gamma_{\nu} \frac{L^{\mu\nu}(k-q)}{(k-q)^2}   \, .
\label{mun2}
\eea

Note, albeit the first is finite and does not require renormalization, this is still  the strength of the kernel $V$ which is responsible for the  mass generation of quarks in the model \cite{VHP}. The second term in Eq. \ref{munk} is a pure  gauge term, i.e.   we neglect the momentum dependence of the vertex in our approximation.

While the observable, like meson masses and cross sections, must be gauge fixing independent,
 the analytical properties of Green's functions do depend on the gauge fixing.  
Facing the necessity of truncation of infinity system of Dyson-Schwinger equations, 
the proper choice of gauge fixing has nontrivial influence on convergence of numerical solution.  
Due to its own importance, a  new procedure of solving  the gap equation (\ref{gap}) 
working for a broad range of gauge fixing parameter $\xi$ is described in the rest of this Appendix.

As the first step, the dimensional regularization was used to express  the selfenergy  as a sum 
 $\Sigma(p,\mu)=\not p \Sigma(p^2,\mu)+\Sigma_B(p^2,\mu)$, 
 each scalar function $\Sigma_A$ and $\Sigma_B$ is then  represented by relatively simple integral over  the 
 spectral quark functions $\sigma_v$ and $\sigma_s$ , which  appear in the representation 
for  the quark propagator: 
\bea  \label{cubik}
S(k)=\int_{\Gamma} d x \frac{\rho(x)}{(\not p-x)} =\int_0^{\infty}d a\frac{\not p \sigma_v(a)+\sigma_s(a)}{p^2-a+\ep} \, .
\eea   
where the later  integral is taken  over  the real  positive domain. 

For a small coupling one can find the solution by solving  the system of equations  numerically   in  manner  described for instance in the review \cite{ja}.
The obtained solution exhibit a presence of a delta function in the spectrum in this soft coupling limit,
 which corresponds to  on shell mass and proper definition of the propagator residuum at that point.   However, when  coupling increases,   the residuum vanishes and  the  resulting system of equations  provide   the only  stable  symmetry preserving solution: $\sigma_{v,s}=0$, in which   we are not interested here.  The nontrivial chiral symmetry breaking 
familiar from many DSEs studies does not occur  within the use of  numerical search \cite{ja}.

It turns out,  it is very useful to  add zero in an identical form, i.e.  subtract the  equation for $A$ and $B$ functions with itself at some arbitrary timelike scale $\zeta$. Thus  one gets
\bea \label{fuc}
B(p^2)&=& B(\zeta) +\Sigma_B(p^2)-\Sigma_B(\zeta) \, ,
\nn \\
 A(p^2)&=&A(\zeta)-\Sigma_A(p^2)+\Sigma_A(\zeta)\, .
\eea
and only after this crucial step one  compares with the assumed  integral  representation (\ref{cubik}).
With the additional assumption that the only singularity of the  propagator is a cut  at the entire  timelike real axis of momenta,  one can write down  the desired set of equations for spectral functions:
\be \label{zdar}
\sigma_s(s)=-\frac{1}{\pi}\frac{\Im B(s) R_D(s)+\Re B(s) I_D(s)}{R_D^2(s)+I_D^2(s)}    \, ,
\ee
where $s=p^2>0$ in our metric, and where the functions  $R_D$ and $I_D$  stand for  the square of the real and the imaginary part of the function $sA^2(s)-B^2(s)$, i.e. 
\bea
R_D(s)&=&s[\Re A(s)]^2-s [\Im A(s)]^2-\Re B(s)^2+[\Im B(s)]^2
\nn \\
I_D(s)&=&2 s \Re A(s)\Im A(s) +2 \Re B(s) \Im B(s) \, ,
 \eea
Similarly for the function $\sigma_v$ one gets
\be \label{zdar2}
\sigma_v(s)=-\frac{1}{\pi}\frac{\Im A(s) I_D(s)-\Im A(s)) R_D(s)}{R_D^2(s)+I_D^2(s)} \, .   
\ee

 First let us note, that  since the equations (\ref{fuc}) represent the expressions for the propagator renormalized in Momentum Renormalization Scheme,  the functions $A$ and $B$ do not depend on the renormalization scale $\mu$ anymore, 
and as usually,  the regularization scheme can be  forgotten  from this point.
However,  since we were subtracting at timelike momentum values rather then at spacelike one, the functions $A(\zeta),B(\zeta)$ $(S_v(\zeta),S_s(\zeta))$ are complex valued and while their real parts can be regarded as a renormalization condition, the angles defined  as
 $\theta_{s,v}=\Re S_{s,v}(x)/ \Im S_{s,v}(x)$ are unique for 
 fixed values of three numbers: $x,\Re A(x),\Re B(x)$. The angles $\theta_{s,v}$ are uniquely determined by  Eq. (\ref{zdar}) noting that $\Im \Sigma_{s,v}(p^2)=-\pi\sigma_{s,v}$ and 
\be \label{condik}
\Re S_{s,v}(s)=P.  \int_0^{\infty}d a\frac{\sigma_{s,v}(a)}{p^2-a}
\ee

It is a matter of  fact, that the equations (\ref{zdar}) and  (\ref{zdar2}) provide numerically convergent solution even for very incorrectly imposed values of angles  $\theta_{s,v}(\zeta)$. To get rid of infinite number of fake solutions,  one needs to implement  the condition (\ref{condik}) into the numerical search from very beginning.

In order to get the solution, we stay with method of iterations and  in addition of making some initial guess of continuous functions $\sigma_{v,s}$,
 we choose also some initial  values for four numbers  $\Re A(\zeta),\Im A(\zeta) $ and $\Re B(\zeta),\Im B(\zeta)$ and solve the system by iterations. 
 If  the system of equations (\ref{zdar}),(\ref{zdar2}) is numerically convergent, then one  change gradually the imaginary parts $\Im A(\zeta) $ and $\Im B(\zeta)$ and look for  a new solutions with desired analyticity improved.

For this purpose  we also construct the auxiliary functions 
\bea \label{cond1}
L_s&=&P.  \int_0^{\infty}d a\frac{\sigma_s(a)}{p^2-a}
\nn \\
R_s&=&\frac{-\Im B(s) I_D(s)+\Re B(s) R_D(s)}{R_D^2(s)+I_D^2(s)}  \, ,       
\eea  
as well as we construct another pair of functions:

\bea \label{cond2}
L_v&=&P.  \int_0^{\infty}d a\frac{\sigma_v(a)}{p^2-a}
\nn \\
R_v&=&\frac{-\Re A(s) R_D(s)+\Im A(s) I_D(s)}{R_D^2(s)+I_D^2(s)}  \, .        
\eea  
The final  search for functions $\Im A(\zeta)$ and $  \Im B(\zeta)$  must yield  
\be \label{equiv2}
L_v(s)=R_v(s)=\frac{1}{4p^2} \Re Tr \not p  S(s)=\Re S_v(s) \, .     
\ee
Therefore we check the  equalities:
 \be \label{equiv1}
L_{s,v}(s)=R_{s,v}(s)     
\ee
 by evaluation the following error norms
\bea
h(\sigma_v,\Im A)&=&\frac{\int (L_v(s)-R_v(s))^2 ds}{\int (L_v(s)+R_v(s))^2 ds} \, ,
\nn \\
h(\sigma_s,\Im B)&=&\frac{\int (L_s(s)-R_s(s))^2 ds}{\int (L_s(s)+R_s(s))^2 ds} \, ,
\label{norms}
\eea
for a given choice of angles $\theta_{v,s}(\zeta)$ at each individual step of  iteration cycle.  Then the functions $\sigma_{v}$ and $\sigma_{s}$, which
are extracted  during the iteration process, turns to be true only in the limit $h\rightarrow 0$. 
 In order to control the numeric, in addition to Equations (\ref{cond1}),(\ref{cond2}),(\ref{equiv1}), another error functions were used.
 They  determine the quality of iterations itself.   
 
Thus there are basically  two iteration cycles, the first, the external one, is used to find a correct value of  angles $\theta_{v,s}(\zeta)$, the second cycle, say the inner one iterates the solution for a fixed angles. The later must provides  vanishing error and   
for a reasonably value of gauge fixing parameter,  only ($\simeq 30$) inner iterations are needed to  provide the stable solution. 
This is  necessary for meaningful evaluation of the errors (\ref{norms}) and  correct value of angles $\theta_{s,v}$  can be identified with reasonable  precision.  In our case, a typical value $h  \simeq 10^{-3} $ is achieved for  hundred of integration mesh points, while  for instance  
  $h \simeq 5. 10^{-5}$ for 4000 integration points.  

\begin{figure}
\centerline{\includegraphics[width=8.6cm]{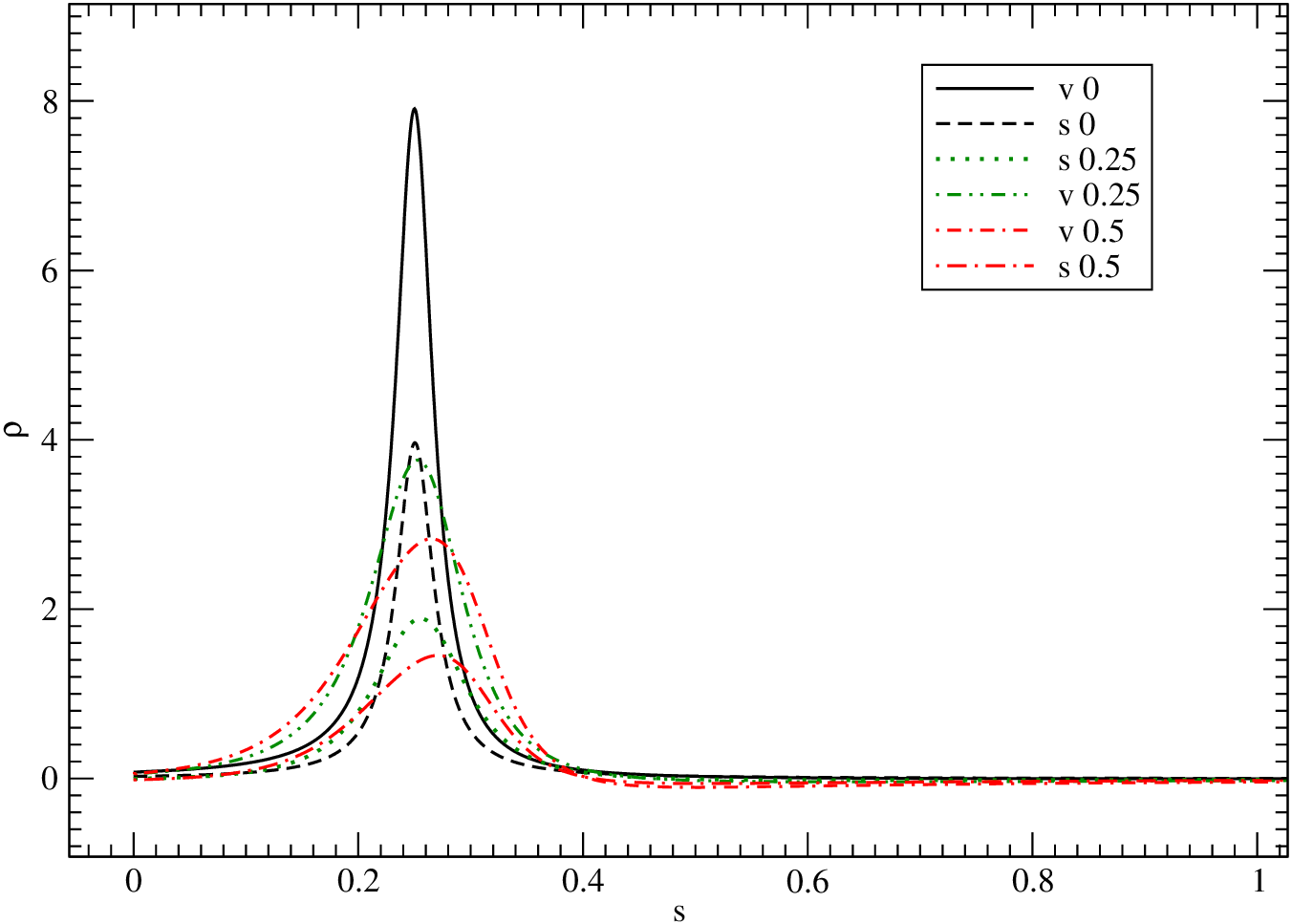}}
\caption{Spectral functions $\sigma_v$ and $\sigma_{s}$ shown for various parameters $(g\xi)^2/(4\pi)^2)$ as described in the Section IV .}
\label{rho2}
{\mbox{-------------------------------------------------------------------------------------}}
{\mbox{-------------------------------------------------------------------------------------}}
\end{figure}

\begin{figure}
\centerline{\includegraphics[width=8.6cm]{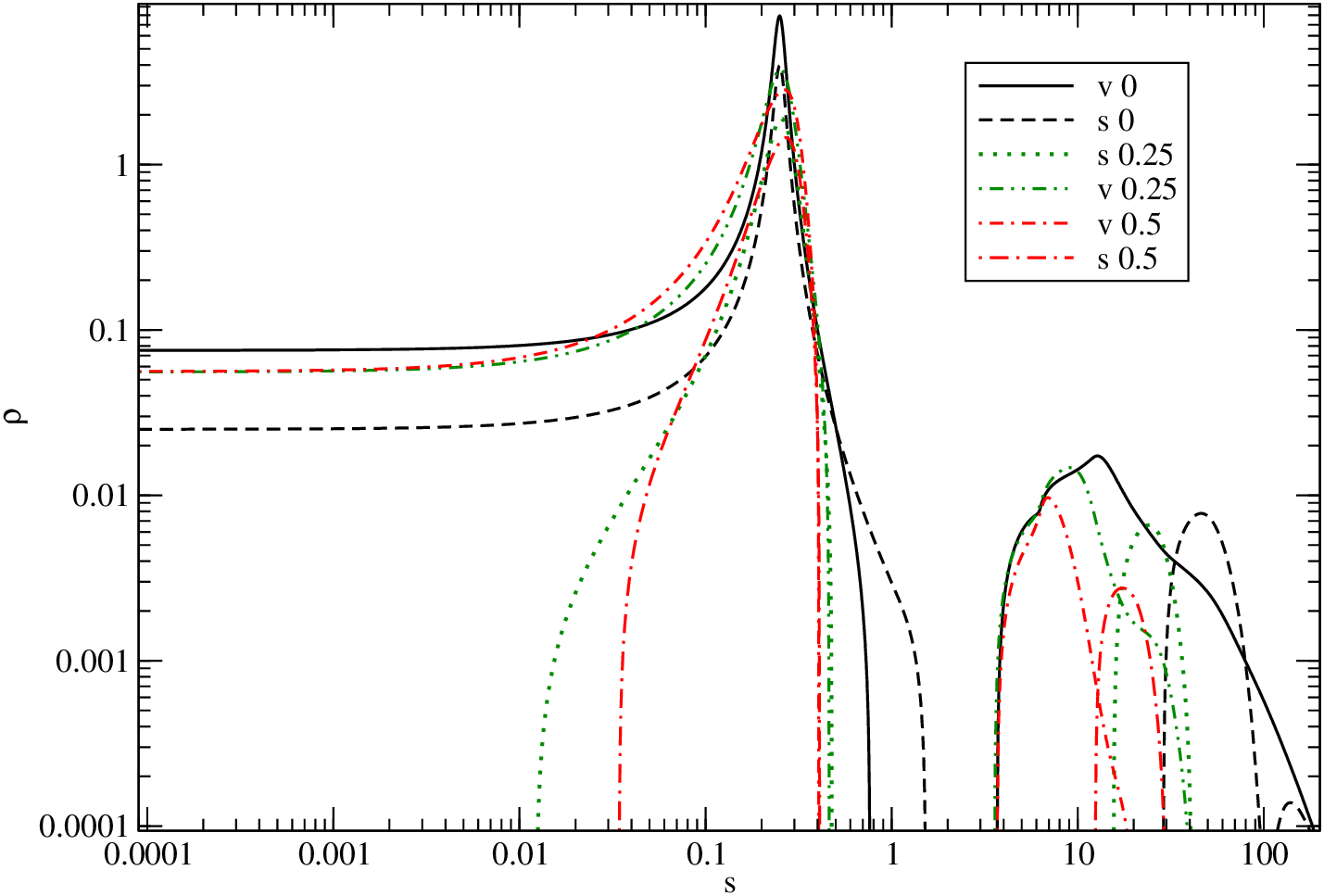}}
\caption{The same as in the previous figure.}
\label{rho}
{\mbox{-------------------------------------------------------------------------------------}}
\end{figure}

Just  for explanatory purpose of this Appendix we do not require the  propagator in (\ref{gap}) to  provide correct physical  
pion properties and we  vary the parameter $\xi$ alone. For clarity let us note, that 
the  effective parameters of our model in Eq. (\ref{gap}) do depend on the gauge fixing parameter 
implicitly (since the DSE kernels are dressed n-point functions).
 Therefore  numerical results presented here should be regarded 
 as different toy  models, rather then single model at different gauges. 
 The mass was fixed such that $B(\zeta=0.5)=m_g$ 
 and $A(\zeta=0.5)=2$ at fixed timelike scale $\zeta=0.5 m_g^2$ and
 $m_g=1$, $m_L=2$, $V_v/(4\pi)^2)=-2$  (presented in units where $m_g=1$). Solutions of the  gap equation for spectral 
 functions $\sigma_v$ (v) and $\sigma_s$ (s)   are shown  for various parameters
$(g\xi)^2/(4\pi)^2=10^{-8},0.25; 0.5 $ in  Fig. \ref{rho2}  and in  Fig. \ref{rho}.

The reader can find similarity  between the procedure described above and the momentum renormalization scheme applied at timelike scale.  
 Likely it can be used in such extent, however stress here, that the method adopted here is not applied  for purpose of  removing UV infinities.
 The renormalization constants stay real, as they were identified by imposing  renormalization conditions 
 at the spacelike t'Hooft renormalization scale  $\mu$.


According to the hypothesis of the quark confinement,  spectral functions
violate the axiom of positivity.  Furthermore,  spectral functions share other important property-
they  do not have thresholds and they do not contain  on mass shell delta function peak in the spectrum. 
The absence of the production threshold in the propagator is  
 consistent with the Wilsonian area law \cite{Wil1974}  which in principle could tell us an
 equivalent information about presence of confinement
without  studying  particular correlator function.

The   absence of a delta function in the spectrum is an obvious and striking  fact and it is equivalent to  the absence 
 of particle pole in the propagator. It automatically  implies  the absence of free moving quarks in the spectrum. 
 We can see a certain  peak formation in both figures \ref{rho} and \ref{rho2}, 
 probably best pronounced for Landau gauge and  less pronounced
 for larger $\xi$ gauges. Thus further questions arise e.g.  what qualitatively specify the confinement,  
 what kind of behavior can be identified from the spectrum  and how the spectrum changes in deconfinement 
 transitions \cite{QR2013,Mas2013,IPRT2018,DORS2019}.


%
\end{document}